\documentclass[aps,twocolumn,showpacs,floatfix,amssymb,preprintnumbers,superscriptaddress]{revtex4}
\usepackage{graphicx}% Include figure files [dvips]
\usepackage{dcolumn}% Align table columns on decimal point
\usepackage{bm}% bold math
\usepackage[dvips]{epsfig}
\usepackage{amssymb}
\usepackage{latexsym}

\begin{document}

\title{Towards a
relevant set of state variables to describe static granular packings}
%\classification{81.05.Rm, 81.20.Ev, 83.80.Fg}
%\keywords {}

\author{Luis A. Pugnaloni}
\email{luis@iflysib.unlp.edu.ar}
\affiliation{Instituto de F\'{\i}sica de L\'{\i}quidos y Sistemas Biol\'{o}gicos (CONICET La Plata, UNLP), cc. 565, 1900 La Plata, Argentina.}
\author{Iv\'{a}n S\'{a}nchez}
\affiliation{Departamento de F\'{\i}sica y Matem\'{a}tica Aplicada, Facultad de Ciencias, Universidad de Navarra, Irunlarrea S/N, 31080 Pamplona, Spain.}
\affiliation{Centro de F\'{\i}sica, Instituto Venezolano de Investigaciones Cient\'{\i}ficas, Apartado Postal 21827, 1020-A Caracas, Venezuela.}
\author{Paula A. Gago}
\affiliation{Instituto de F\'{\i}sica de L\'{\i}quidos y Sistemas Biol\'{o}gicos (CONICET La Plata, UNLP), cc. 565, 1900 La Plata, Argentina.}
\author{Jos\'{e} Damas}
\affiliation{Departamento de F\'{\i}sica y Matem\'{a}tica Aplicada, Facultad de Ciencias, Universidad de Navarra, Irunlarrea S/N, 31080 Pamplona, Spain.}
\author{Iker Zuriguel}
\affiliation{Departamento de F\'{\i}sica y Matem\'{a}tica Aplicada, Facultad de Ciencias, Universidad de Navarra, Irunlarrea S/N, 31080 Pamplona, Spain.}
\author{Diego Maza}
\email{dmaza@unav.es}
\affiliation{Departamento de F\'{\i}sica y Matem\'{a}tica Aplicada, Facultad de Ciencias, Universidad de Navarra, Irunlarrea S/N, 31080 Pamplona, Spain.}

\begin{abstract}
We analyze, experimentally and numerically, the steady states, obtained by tapping, of a 2D granular layer. Contrary to the usual assumption, we show that the reversible (steady state branch) of the density--acceleration curve is nonmonotonous. Accordingly, steady states with the same mean volume can be reached by tapping the system with very different intensities. Simulations of dissipative frictional disks show that equal volume steady states have different values of the force moment tensor. Additionally, we find that steady states of equal stress can be obtained by changing the duration of the taps; however, these states present distinct mean volumes. These results confirm previous speculations that the volume and the force moment tensor are both needed to describe univocally equilibrium states in static granular assemblies.
\end{abstract}

\maketitle

\emph{Introduction:} Finding out the appropriate set of macroscopic variables that characterizes the equilibrium state of a given system is the first step in any thermodynamic study \cite{callen}. What these variables are for a sample of sand in a box is still under discussion \cite{leiden}.
Twenty years ago, Edwards and Oakshott\cite{edwards} put forward the idea that the number of grains $N$ and the volume $V$ are the basic state variables that suffice to characterize a static granular sample in equilibrium. The $NV$ granular ensemble was introduced as a collection of microstates, where the sample is in mechanical equilibrium, compatible with $N$ and $V$.

Experimentally, such equilibrium states are commonly obtained by tapping the system at different dimensionless accelerations, $\Gamma$, which is the standard parameter used to quantify the intensity of the external excitation \cite{nowak,nowak2,richard,schroter,ribiere}. Hence, under the action of repeated excitations at constant $\Gamma$, the system properties reach a steady state where all observables fluctuate around well defined mean values. Therefore, tapping provides the energy necessary to allow the system to explore different configurations over which macroscopic observables can be averaged. In the steady state, the mean volume $V$, or, more commonly, the mean packing fraction, $\phi$ (defined as the percentage of the space occupied by the particles), can be measured. It is now well established \cite{nowak,nowak2,richard,schroter,ribiere} that different steady states can be reached reversibly by varying $\Gamma$. These experiments also suggest a monotonic relation between $\Gamma$ and $\phi$. Nevertheless, $\phi$ does not only depend on $\Gamma$. In harmonic pulses, if $\Gamma$ is kept constant while the duration and the amplitude of the pulse are simultaneously modified, the system evolves to different steady states. This fact was used to show that equivalent equilibrium states \cite{pica,schroter} can be generated by using different combinations of pulse amplitude and duration. On the other hand, new theoretical works \cite{snoeijer,edwards2,blumenfeld,henkes,henkes2,tighe} suggest that the force moment tensor, $\Sigma$, ($\Sigma=V \sigma$, where $\sigma$ is the stress tensor) must be added to the set of extensive macroscopic variables (i.e., an $NV\Sigma$ ensemble) to describe adequately a packing of grains.

 Recent numerical simulations \cite{pugnaloni,gago} have predicted also the existence of a nonmonotonic relation between $\Gamma$ and $\phi$, if sufficiently large values of $\Gamma$ are considered. A hint as to this behavior can already be seen in some packings obtained in the laboratory \cite{sibille};  although this has passed unnoticed by those studying equilibrium properties. These results imply that equilibrium states of equal $V$ (or $\phi$) could be obtained not only by changing the pulse duration, as in Ref. \cite{schroter}, but also by simply using different values of the excitation parameter $\Gamma$. In what follows, we will show that these equal volume equilibrium states are experimentally accessible. We will then use numerical simulations to show that these states are, nevertheless, different. We have found that two equilibrium states obtained with different excitations, even if they have same $V$ and $N$, may present distinct values of $\Sigma$. This result suggests the incomplete picture upon which research has been based for over two decades when the properties of static granular matter have been investigated; and takes us one step closer to completing the definition of the state variables that are necessary to describe a static packing of grains.

\begin{figure}[t]
\includegraphics[width=1\columnwidth,trim=4mm 4mm 6mm 5mm, clip]{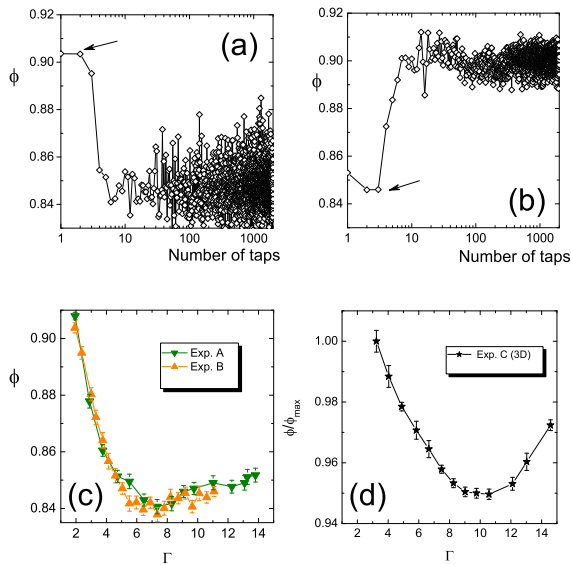}
\caption {(Color online)(a) Evolution of
$\phi$ towards the steady state at $\Gamma = 7.0$ starting from the steady state
corresponding to $\Gamma = 1.5$. (b) The evolution when switching back to $\Gamma = 1.5$ \cite{movie1} (Arrows indicate the tap where $\Gamma$ is switched). (c) Steady state packing fraction, $\phi$, as a function of $\Gamma$ (green down triangles: steady state obtained in Exp A through tapping at
constant $\Gamma$ from an initial ordered state; orange up triangles: steady state in Exp B obtained by
an annealing protocol) \cite{nowak,movie1}. (d) Steady state relative packing
fraction $\phi / \phi_{\rm{max}}$ as a function of tapping intensity $\Gamma$ for the 3D cell ($\phi_{\rm{max}}$ corresponds to the value of $\phi$ obtained at the lowest $\Gamma$ studied). Error bars represent the standard error
of the mean estimated from the standard deviation.}.
\label{figure1}
\end{figure}

\emph{Experimental results:} A quasi 2D Plexiglass cell (width: 28 mm, height: 150 mm) was used to study
the packing dynamics. The cell was filled with 1000 alumina oxide (Al$_2$O$_3$) spheres of diameter
1.000 $\pm$ 0.005 mm. The separation between the Plexiglass sheets was 10\% larger than the bead diameter in order to
minimize the particle-wall friction and prevent arching in the transversal direction. The system
was tapped with an electromagnetic shaker (Tiravib 52100) with a series of harmonic pulses of
variable amplitude and constant frequency $\nu = 30$ Hz every three seconds. The tapping intensity was measured with a piezoelectric accelerometer attached to the base of the cell. Although in recent years alternative parameters have been proposed to quantify the intensity of the taps \cite{dijksman,ludewing}, we employ the usual nondimensional peak acceleration $\Gamma \equiv a_{peak}/g$ (where $g$ is the acceleration of gravity). High resolution digital images of the packing were taken after each tap. The center of each sphere was detected with an error of less than 2\%. The packing fraction of each image was obtained by considering each grain as a disk of
the corresponding effective diameter and then calculating the percentage of the area covered by
the disks in a rectangular area 10\% smaller than the size of the packing. Due to the space left
between the Plexiglass sheets, small overlaps can be observed in the 2D projection taken by the
photographs. This effect leads to estimations of $\phi$ above the 2D hexagonal close packing value
for the densest configurations studied. After a relative short number of taps, the system reaches a steady state characterized by
a plateau in curves of $\phi$ vs. the number of taps. In contrast with the three dimensional case, any stationary steady state is obtained after a few taps. We checked that the mean value and standard deviation of the packing fraction differ less than $0.1\,\%$ if $200$ or $2000$ taps of equilibration are applied. As an example, in Fig. 1(a), we show that the steady state corresponding to $\Gamma = 7.0$ can be reached in a couple taps, even when the initial packing is in an ordered configuration. Equilibrating the system back at $\Gamma = 1.5$ takes somewhat longer (about 100 taps) [see Fig. 1(b)] \cite{movie1}.
Therefore, the sample initially prepared in a
highly ordered configuration was tapped 500 times for any given $\Gamma$ before taking averages over 100 taps, and $10$ independent runs were averaged. Alternatively, an annealing protocol was used in which $\Gamma$ was decreased in discrete steps from
the highest values and tapped 500 times at each $\Gamma$ value without emptying the cell.

In Fig. 1(c), we plot $\phi$ in the steady state as a function of $\Gamma$ for different independent repetitions of the experiment. It can be seen that the same results are obtained by equilibrating from an initial ordered structure (Exp. A) or by following an annealing path (Exp. B). Although the fluctuations of $\phi$ are large [see Fig, 1(a)], its mean value is well defined with a small confidence interval [see error bars in Fig. 1(c)]. For low excitations, $\phi$ decreases as $\Gamma$ is increased, in agreement with previous results reported by several groups. However, beyond a certain value $\Gamma_{\rm{min}}$, the packing fraction grows \cite{movie1}. The same trend is observed if the tap frequency $\nu$ is changed (Exp. C) [see Fig 2(a)].  An explanation for this behavior based on the formation of arches has been given in \cite{pugnaloni}.

%
%\begin{figure}[t]
%\includegraphics[width=0.9\columnwidth,trim=4mm 4mm 6mm 5mm, clip]{figure2}
%\caption {(a) Steady state packing fraction, $\phi$, as a function of $\Gamma$ when
%tapping at $\nu = 30$ Hz (green down triangles: steady state obtained in Exp A through tapping at
%constant $\Gamma$ from an initial ordered state; orange up triangles: steady state in Exp B obtained by
%an annealing protocol) \cite{nowak,movie2}. (b) Steady state relative packing
%fraction $\phi / \phi_{\rm{max}}$ as a function of tapping intensity $\Gamma$ for the 3D cell ($\phi_{\rm{max}}$ corresponds to the value of $\phi$ obtained at the lowest $\Gamma$ studied). Error bars represent the standard error
%of the mean estimated from the standard deviation.}
%\label{figure2}
%\end{figure}

\begin{figure*}[t]
\includegraphics[width=1.8\columnwidth,trim=0mm 0mm 0mm 0mm, clip]{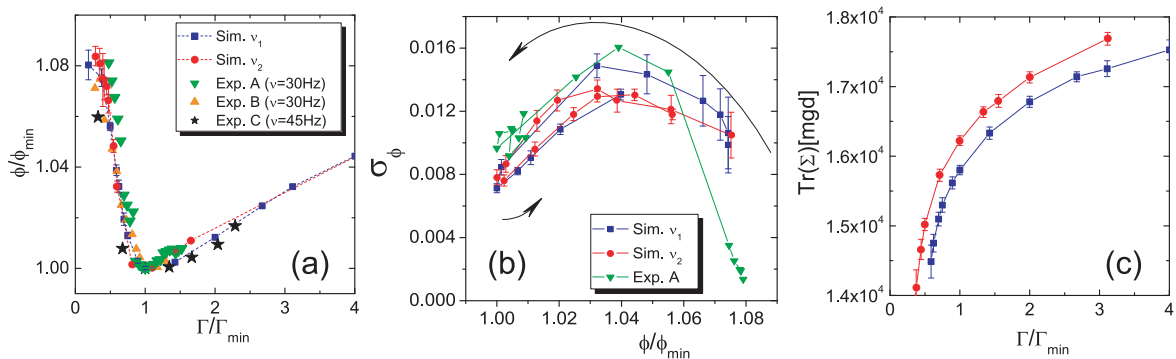}
\caption {(Color online). Results from the MD simulations of the soft disk model for two different
frequencies of the tapping pulse [blue squares: $\nu_1 = 0.5 (g/d)^{1/2}$, and red circles: $\nu_2 = 0.25 (g/d)^{1/2}$]. (a) Steady state packing fraction, $\phi$, as a function of the drive $\Gamma$. In order to compare with
the quasi 2D experiments (green down triangles, orange up triangles and black stars) the vertical and horizontal axes
have been scaled with the characteristic values $\phi_{\rm{min}}$ and $\Gamma_{\rm{min}}$, respectively. (b) Fluctuations, $\sigma_\phi$, of the packing fraction in the steady state
(measured by the standard deviation) as a function of $\phi$. The arrows indicate the direction of increasing tapping intensity. Error bars correspond to the standard error.  (c) The trace, Tr$(\Sigma)$, of the force moment tensor
in the steady state as a function of $\Gamma/\Gamma_{\rm{min}}$.}
\label{fig2}
\end{figure*}

In order to assess whether the presence of a minimum in the $\Gamma$--$\phi$ curves is induced by the highly ordered crystal-like structures present in the quasi 2D cell, we repeated the experiment with a 3D cell. This 3D setup consists of a cell of the same height and width (as the quasi 2D cell) but 6 mm thick. In this case, the granular sample was made of polydisperse glass beads 1.0 $\pm$ 0.2 mm in diameter. For the sake of simplicity, in the 3D cell, the relative packing fraction was estimated from the height of the granular layer. The stationary
regime in the 3D setup was obtained after $2 \times 10^4$ taps. In Fig. 1(d) we show the steady state packing fraction as a function of $\Gamma$ for this three-dimensional setup. Again, the same trend as in the quasi 2D experiment is observed. This nonmonotonic dependence challenges the idea that the volume defines the equilibrium state, unless these equal volume states are proved to be equivalent \cite{mehta}.

\emph{Numerical evidences:} In 2005, Edwards suggested that the stress tensor should be included in the description of the equilibrium state of a static granular sample along with $N$ and $V$ in the so-called full canonical ensemble \cite{edwards2}. There is some consensus now \cite{snoeijer,blumenfeld,henkes,henkes2,tighe} that the force moment tensor, $\Sigma \equiv V\sigma$ , allows for a microcanonical description \cite{henkes2}. However, no experiments or simulations have shown that this variable is necessary to distinguish between different states of thermodynamic equilibrium.

Since it is difficult to measure the force moment tensor in our experimental setup, we use a realistic computer model of the quasi 2D setup \cite{pugnaloni,arevalo}. We used a velocity Verlet algorithm to integrate the Newton equations for 512 monosized disks in a rectangular box. The disk--disk and disk--wall contact interaction comprises a linear
spring--dashpot in the normal direction and a tangential friction force that implements the Coulomb criterion to switch between dynamic and static friction. Units are reduced with the diameter of the disks, $d$, the disk mass, $m$, and the acceleration of gravity, $g$. Details on the force equations and the interaction parameters can be found elsewhere \cite{pugnaloni,arevalo}. Tapping is simulated by moving the confining box in the vertical direction following a half sine wave trajectory. The intensity of the excitation is controlled either through the amplitude, $A$, or the frequency, $\nu$, of the sinusoidal trajectory; and it is characterized by the parameter $\Gamma = A (2\pi\nu)^2/g$. We consider the system has reached the steady state whenever a plateau in
$\phi$ is observed (with no visible trend in a plot of $\phi$ vs
$\log(taps)$). Averages were taken over 400 taps in the steady state and over
20 independent simulations for each value of $A$ and $\nu$.

In Fig. 2(a), we show $\phi$ as a function of $\Gamma$ for this model. Different values of $\Gamma$ were obtained by varying both $A$ and $\nu$. As in the experiment, states of equal $\phi$ are generated at both sides of a minimum packing fraction, $\phi_{\rm{min}}$. States of equal volume at each side of $\phi_{\rm{min}}$ display slightly different  volume fluctuations [Fig. 2(b)] suggesting that these states are not the same. In contrast to the packing fraction, the trace of $\Sigma$ \cite{sigma} gently grows as $\Gamma$ is increased [see Fig. 2(c)]. There is neither a minimum nor a maximum in the $\Gamma$--$\Sigma$ curves. Therefore, states of equal $\phi$ at each side of $\phi_{\rm{min}}$ present distinct $\Sigma$. Hence, states of equal volume at each side of $\Gamma_{\rm{min}}$ are, in fact, different. This finding proves that the volume alone cannot characterize the equilibrium states of a static granular sample and suggests that the force moment tensor is a convenient extra state variable.

\begin{figure}[t]
\includegraphics[width=1\columnwidth,trim=0mm 0mm 0mm 0mm, clip]{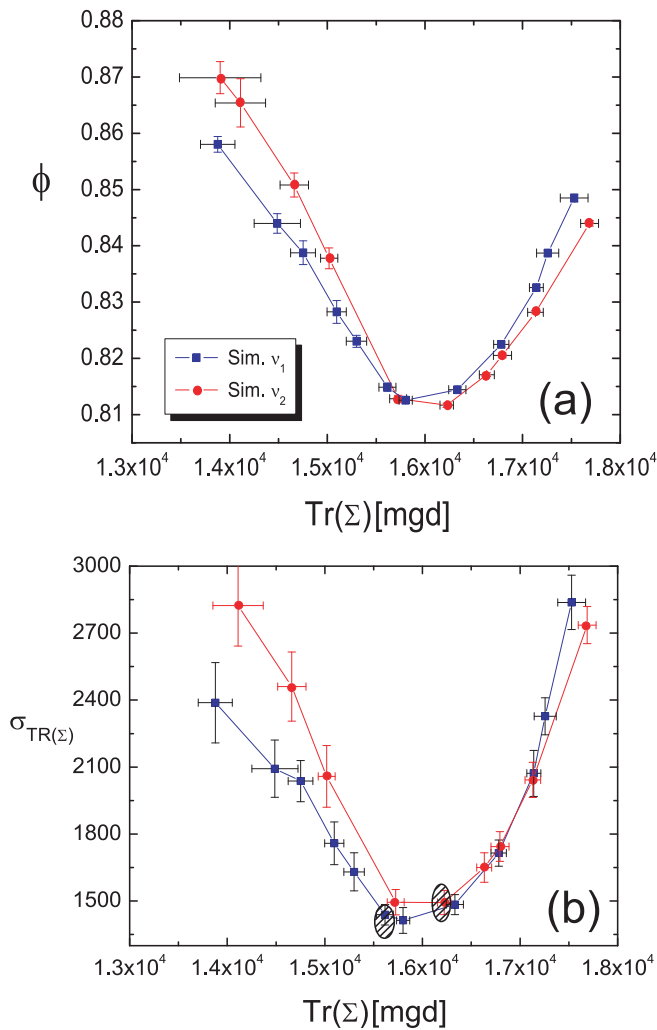}
\caption {(Color online). The loci of the generated equilibrium states in the $\phi$--$\Sigma$ phase diagram.
(a) The stationary packing fraction as a function of the trace Tr$(\Sigma)$ for $\nu_1$ (blue squares) and $\nu_2$ (red circles). (b) The fluctuations, $\sigma_{\rm{Tr}(\Sigma)}$, versus Tr$(\Sigma)$. Error bars as in Fig. 2. The shaded areas indicate the position of $\phi_{\rm{min}}$.
}
\label{fig3}
\end{figure}

In Fig. 3(a), the loci defined by the equilibrium states visited during the numerical tapping experiments are plotted in a hypothetical $\phi$--$\Sigma$ thermodynamic phase space. It is clear that states of equal $\Sigma$ can present different $\phi$ if prepared at different $\nu$. Thus, $\Sigma$ cannot define the equilibrium state by itself. Both $V$ and $\Sigma$ have to be specified to fully identify a given state. This result is confirmed by the fluctuations of the force moment tensor, $\sigma_\Sigma$, [see Fig. 3(b)]. The fluctuations $\sigma_\Sigma$ for a given $\Sigma$ are not unique: they depend on the frequency $\nu$ of the tapping. However, for the states obtained with different excitations that display 
equal $\phi$ and $\Sigma$ (note that such states would actually be a 
unique state), we find that $\sigma_\phi$ and $\sigma_\Sigma$ display 
also very similar values. Hence, the coincidence of the mean 
and fluctuations of the state variables suggests that no other extensive parameter would be 
necessary to describe such equilibrium states.

It is interesting to mention that the definition of an entropy $S(N,V,\Sigma)$ is particularly convenient since $\Sigma$ is not bounded as $V$ is. This implies that $S$ can be a monotonic increasing function of $\Sigma$, which permits a well behaved Legendre transformation to a canonical ensemble \cite{callen}. Hence, one can define a non-negative temperature-like quantity: the angoricity \cite{edwards2,henkes2} ---the proposed name for the inverse of the conjugate variable to the force moment tensor. Let us finally recall that the temperature-like quantity associated with the volume (so-called compactivity) can present negative values due to population inversion \cite{brey}. 

\emph{Conclusions:} We have shown that steady states of static granular packings obtained by tapping the system with different pulse strength and duration can present different mean force moment tensor even if they correspond to the same mean volume. Moreover, steady states that present the same mean force moment tensor are distinguishable by their mean volume. To our knowledge, this is the first experimental/numerical evidence that both extensive variables should be included in an entropic formulation of the thermodynamics of such steady states.
\acknowledgments
This work was supported by project FIS2008-06034-C02-01 (Spain), PIUNA (Univ. Navarra), CONICET (Argentina) and ANPCyT (Argentina).


\begin{thebibliography}{50}
\bibitem{sigma} We have seen that the two invariants of $\Sigma$ (i.e. the trace and the determinant in 2D) are proportional to one another (i.e. $Det(\Sigma)= a Tr(\Sigma) + b$). Moreover $a$ and $b$ are independent of the amplitude and duration of the taps used to reach a given $\Sigma$. Therefore, $Tr(\Sigma)$ suffices to characterize $\Sigma$ for our purposes.
\end{thebibliography}
\end{document}